\documentclass[12pt]{iopart}
\usepackage{iopams}
\usepackage{graphicx}
\usepackage{harvard}
\begin{document}
\title[Self-organized criticality in a neural network]{\bf Self-organized criticality in a network of interacting neurons}

\author{J D Cowan$^1$,  J Neuman$^2$, W van Drongelen$^3$}
\address{$^1$ Dept. of Mathematics, University of Chicago, 5734 S. University Ave., Chicago, IL 60637}
\address{$^2$ Dept. of Physics, University of Chicago, 5720 S. Ellis Ave., Chicago, IL, 60637}
\address{$^3$ Dept. of Pediatrics, University of Chicago, KCBD 900 E. 57th St., Chicago, IL., 60637}
\ead{cowan@math.uchicago.edu}
\begin{abstract}
This paper contains an analysis of a simple neural network that exhibits self-organized criticality.
Such criticality follows from the combination of a simple neural network with an excitatory feedback loop that generates bistability, in combination with an anti-Hebbian synapse in its input pathway. Using the methods of statistical field theory, we show how one can formulate the stochastic dynamics of such a network as the action of a path integral, which we then investigate using renormalization group methods. The results indicate that the network exhibits hysteresis in switching back and forward between its two stable states, each of which loses its stability at a saddle-node bifurcation. The renormalization group analysis shows that the fluctuations in the neighborhood of such bifurcations have the signature of directed percolation. Thus the network states undergo the neural analog of a phase transition in the universality class of directed percolation.  The network replicates precisely the behavior of the original sand-pile model of Bak, Tang \& Wiesenfeld.
\end{abstract}

\maketitle

\section{Introduction}
The idea of {\it self-organized criticality} (SOC) was introduced by~\citeasnoun{Bak88}. Their paper immediately triggered an avalanche of papers on the topic, not the least of which was a connection with $1/f$- or scale-free noise. However it was not until another paper appeared, by~\citeasnoun{Gil96}, which greatly clarified the dynamical prerequisites for achieving SOC, that a real understanding developed of the essential requirements for SOC: (1) an {\it order-parameter} equation for a dynamical system with a time-constant $\tau_o$, with stable states separated by a threshold, (2)  a {\it control-parameter} equation with a time-constant $\tau_c$, and (3) a steady {\it driving force}. In Bak {\it et.al.}'s classic example, the sand-pile model, the order parameter is the rate of flow of sand grains down a sand-pile, the control parameter is the sand-pile's slope, and the driving force is a steady flow of grains of sand onto the top of the pile.  Gil and Sornette showed that if $\tau_o \ll \tau_c$ then the resulting avalanches of sand down the pile would have a scale-free distribution, whereas if $\tau_o \gg \tau_c$ then the distribution would also exhibit one or more large avalanches.

In this paper we will analyze a neural network model which is in one-to-one correspondence with the Gil-Sornette SOC-model, and therefore also exhibits SOC.

\subsection{Neural network dynamics}
Consider first the mathematical representation of the dynamics of a neocortical slab comprising a single spatially homogeneous network of $N$ excitatory neurons. Such neurons make transitions from a quiescent state $q$ to an activated state $a$ at the rate $\sigma$ and back again to the quiescent state $q$ at the rate $\alpha$, as shown in figure 1.
\begin{figure}[thbp]
\begin{center}
\includegraphics[width = 3.9 cm]{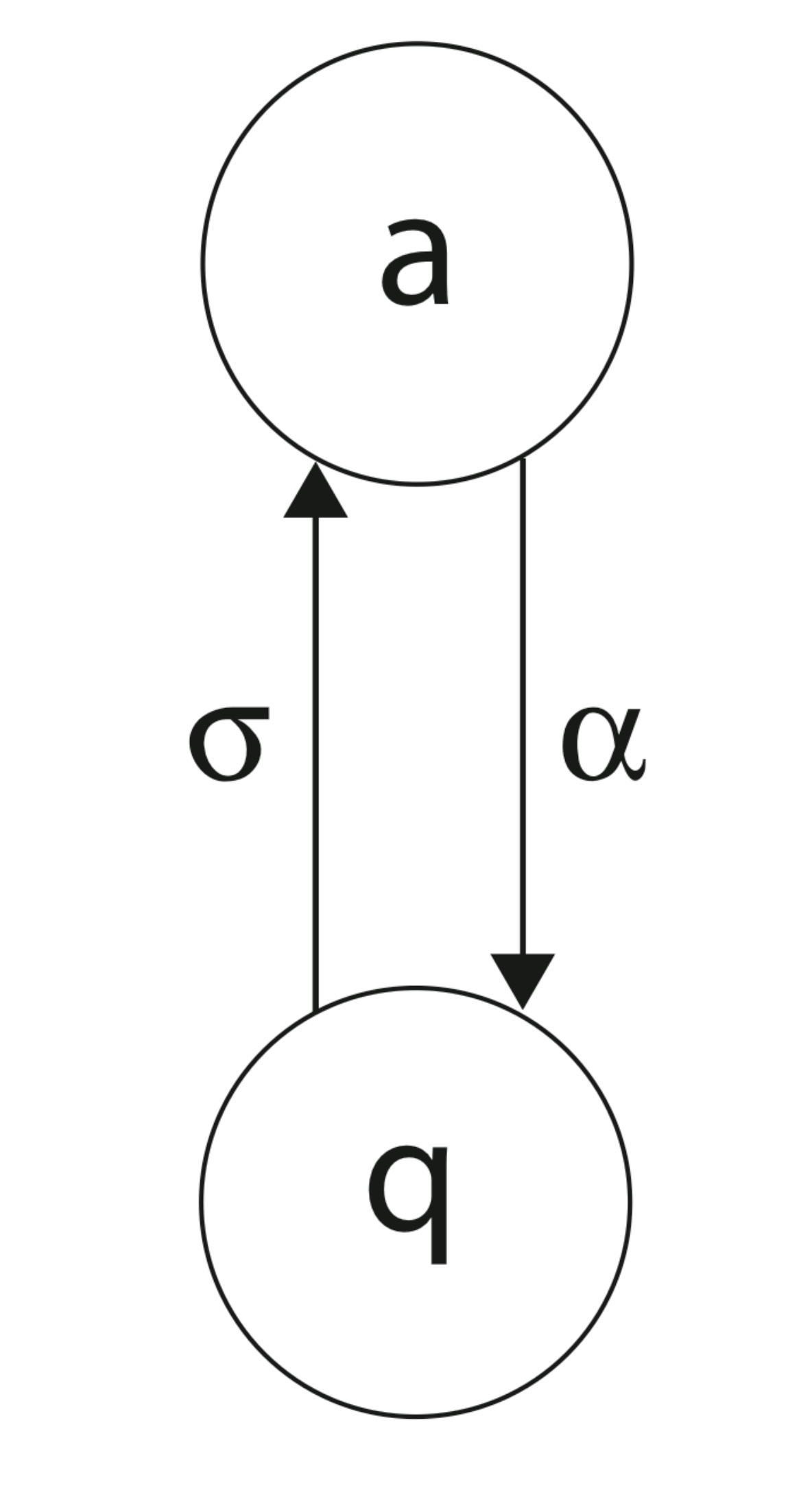}
\caption{\small Neural state transitions}
\end{center}
\label{fig:1}
\end{figure}

\noindent Let $P_{n}(t)$ be the probability that a fraction $n/N$ is {\em active} at time $t$. Then the probabilistic dynamics of such a slab can be formulated as a master equation of the form:
\begin{eqnarray}
\frac{dP_n(t)}{dt}&=&\alpha[(n+1)P_{n+1} - nP_n] 
+(N-n+1)f[s(n-1)]P_{n-1} \nonumber \\
&-&(N-n)f[s(n)]P_n 
\label{eq:ME1}
\end{eqnarray}
where $\alpha$ is the rate at which activated neurons become {\em quiescent}, and $s(n)$ is the total current or excitation driving each neuron in the population to fire at the rate $\sigma = f[s(n)]$.  We assume in this simplified model that each neuron receives a signal weighted by $w_0/N$ from each of   $N$ other neurons in the population,  so that 
\begin{equation}
s(n) = w_0 n + h
\label{eq:C1}
\end{equation} 
where $h$ is an external current.  

Equation~\ref{eq:ME1} can be extended to the spatially inhomogeneous case.  Let $n_r/N_r$ be the fraction of active cells at time $t$ in the $r$th population of $N_r$ cells, and let $P[\mathbf{n},t]$ be the probability of the configuration $\mathbf{n} = \{n_1, n_2, \cdots, n_r, \cdots, n_\Omega\}$ existing at time $t$.  The extended master equation then takes the form
\begin{eqnarray}
\frac{dP[\mathbf{n},t]}{dt}&=&\alpha \sum_r\left[(n_r+1)P[\mathbf{n}_{r+},t] - n_rP[\mathbf{n},t]\right] \nonumber \\
&+&\sum_r\left[(N_r-n_r +1)f[s(n_r -1)]P[\mathbf{n}_{r-},t] \right. \nonumber \\
 &-& \left.(N_r-n_r)f[s(n_r)]P[\mathbf{n},t]\right] 
\label{eq:ME2}
\end{eqnarray}
where $\mathbf{n}_{r \pm}=\{n_1,n_2, \cdots,n_{r} \pm 1, \cdots, n_\Omega\}$ and there are a total of $\Omega$ locally homogeneous populations.

Following~\citeasnoun{VanKamp81}  equation~\ref{eq:ME2} can be rewritten using the one-step operators
\begin{equation}
\mathcal{E}^{\pm}_r f(\mathbf{n})= f(\mathbf{n}_{r }\pm 1)
\label{eq:OS1}
\end{equation}
i.e.,
\begin{equation}
\frac{dP[\mathbf{n},t]}{dt}= \sum_r  \left[\alpha( \mathcal{E}^+_r -1)n_r
+ ( \mathcal{E}^-_r -1)(N_r-n_r)f[s(n_r)]\right]  P[\mathbf{n},t] 
\label{eq:ME3}
\end{equation}

Note that the total number of cells $N_r$ in the $r$th population comprises $n_r$ active cells and $N_r-n_r = q_r$ quiescent cells, so that 
\begin{equation}
n_r + q_r = N_r
\label{eq:Sum}
\end{equation} 
Thus equation~\ref{eq:ME3} can be rewritten slightly in the form
\begin{equation}
\frac{dP[\mathbf{n},t]}{dt}= \sum_r  \left[\alpha( \mathcal{E}^+_r -1)n_r
+ ( \mathcal{E}^-_r -1)q_r f[s(n_r)]\right]  P[\mathbf{n},t] 
\label{eq:ME4}
\end{equation}

\subsection{Annihilation and creation operators}
We now introduce Fock space {\it annihilation} and {\it creation} operators satisfying boson commutation rules
\begin{eqnarray}
	&&[a_r, a^{\dag}_s] = [q_r, q^{\dag}_s] = \delta(r-s) \nonumber \\
          &&[a_r, a_s] = [a^{\dag}_r, a{\dag}_s] = 0 \nonumber \\
          &&[q_r, q_s] = [q^{\dag}_r, q^{\dag}_s] = 0  
\label{eq:Boson1}
\end{eqnarray}
such that given neural vectors $| n_r \rangle $, $ | m_{r'} \rangle $
\begin{eqnarray}
a_r^\dag | n_r \rangle & = & | n_r+1  \rangle, a_r | n_r \rangle = n_r | n_r-1  \rangle \nonumber \\
q_r^\dag | m_r \rangle & = & | m_r+1  \rangle, q_r | m_r \rangle = m_r | m_r-1  \rangle 
\label{eq:Boson 2}
\end{eqnarray}
\noindent where 
\begin{equation}
n_r+m_r=1, | n_r \rangle = (a^\dag)^n_r | 0_r \rangle, | m_r \rangle = (q_r^\dag)^m_r | 0_r \rangle
\label{eq:Boson3}
\end{equation}
\noindent and $|0_r \rangle$ is a fiducial or vacuum (empty) state such that
\begin{eqnarray}
a_r^\dag |0_r \rangle = |n_r=1_r \rangle, a_r |0_r \rangle=0  \nonumber \\
q_r^\dag |0_r \rangle = |m_r=1_r \rangle, q_r |0_r \rangle=0  \label{eq:Boson4}
\end{eqnarray}
\noindent and there exists a {\it dual} vector $\langle 0 |$ such that
\begin{eqnarray}
\langle 0_r | a_r = \langle n_r=1_r |, \langle 0_r | a_r ^\dag = 0, \nonumber \\
\langle 0_r | q_r = \langle m_r=1_r |, \langle 0_r | q^\dag_r = 0, \label{eq: Boson5}
\end{eqnarray}

\noindent It follows from equation~\ref{eq:Boson 2} that $$a^\dag_r a_r | n_r \rangle = n_r | n_r \rangle$$ i.e. $n_r$ is the {\it eigenvalue} of the operator $a^\dag_r a_r$, which is therefore referred to as a {\it number} or {\it number density} operator.

\subsection{Bosonizing the master equation}
The master equation can now be {\it bosonized} by replacing the van Kampen one-step operators by bosonic equivalents, i.e.
\begin{eqnarray}
(\mathcal{E}^+_r - 1)  &=& (q_r^\dag - a_r^\dag) a_r  \nonumber \\
(\mathcal{E}^-_r - 1)  &=& (a_r^\dag - q_r^\dag) q_r 
\label{eq: Boson6}
\end{eqnarray}

\noindent so eqn~\ref{eq:ME4} becomes:
\begin{equation}
\frac{dP[\mathbf{n},t]}{dt}= \sum_r  \left[\alpha(q_r^\dag - a_r^\dag) a_r
+  (a^\dag_r - q_r^\dag) q_r f[s(a^\dag_r a_r)]\right]  P[\mathbf{n},t] 
\label{eq:ME5}
\end{equation}

\noindent Let $|P(t)\rangle$ be a probability state vector satisfying
\begin{equation}
|P(t)\rangle = \sum_{\mathbf{n}} P[\mathbf{n},t] |\mathbf{n}\rangle
\label{eq:ME6}
\end{equation}
\noindent Then equation~\ref{eq:ME4} can be written in the form
\begin{equation}
\frac{d}{dt}|P(t)\rangle=\sum_r \left[\alpha(q_r^\dag - a_r^\dag) a_r
+  (a^\dag_r - q_r^\dag) q_r f[s(a^\dag_r a_r)]\right] |P(t)\rangle
\label{eq:ME7}
\end{equation}
\noindent or formally as
\begin{equation}
\frac{d}{dt}|P(t)\rangle=-\hat{H} |P(t)\rangle
\label{eq:ME8}
\end{equation}
where
\begin{equation}
-\hat{H}=\sum_r \left[\alpha(q_r^\dag - a_r^\dag) a_r
+  (a^\dag_r - q_r^\dag) q_r f[s(a^\dag_r a_r)]\right]
\label{eq:QHam}
\end{equation}
\noindent is the {\it quasi-Hamiltonian} operator for the Markov process represented in equation~\ref{eq:ME1}.

\subsection{From bosons to coherent states}
Equation~\ref{eq:QHam} is a linear operator equation with formal solution $$|P(t)\rangle = \exp[-\hat{H}(t-t_0)|P(t_0)\rangle$$ We need to re-express this solution in terms of  numbers rather than operators. This can be achieved by introducing {\it coherent states}. These were introduced by ~\citeasnoun{Schro26} and first used extensively in coherent optics by~\citeasnoun{Gla63}. We therefore introduce such states $|\phi_r \rangle$ in the form
\begin{equation}
|\phi_r \rangle = \exp[-\frac{1}{2} \varphi_r^\star \varphi_r + \varphi_r a^\dag_r ] | 0_r \rangle
\label{eq:CS1}
\end{equation}
where $\varphi_r$ is the right eigenvalue of $a_r$, i.e. $a_r | \phi_r \rangle = \varphi_r | \phi_r \rangle$.
There is also a coherent state representation of $q_r$ in the form $|\theta_r \rangle$ such that the right eigenvalue of $q_r$ is $\vartheta_r$, i.e. $q_r |\theta_r \rangle = \vartheta_r |\theta_r \rangle$.  In similar fashion $\langle \phi_r | a^\dag_r = \langle \phi_r | \tilde{\varphi}_r$ where $\tilde{\varphi}_r$, the complex conjugate of $\varphi$, is the left eigenvalue of $a^\dag_r$, and similarly $\langle \theta_r | q^\dag_r = \langle \theta_r |\tilde{\vartheta}_r$, i.e. $\tilde{\vartheta}_r$ is the left eigenvalue of $q^\dag_r$.  It follows  that
\begin{equation}
\langle \phi_r | a^\dag_r a_r | \phi_r \rangle = \langle \phi_r | \tilde{\varphi}_r \varphi_r | \phi_r \rangle = \tilde{\varphi}_r \varphi_r 
\label{eq:CS2}
\end{equation}
\noindent All this suggests that the operator quasi-Hamiltonian has a coherent state representation in the form
\begin{equation}
-\mathcal{H} = \sum_r \left[ \alpha(\tilde{\vartheta}_r - \tilde{\varphi}_r ) \varphi_r
+  (\tilde{\varphi}_r - \tilde{\vartheta}_r ) \vartheta_r  f[s(\tilde{\varphi}_r \varphi_r)] \right]
\label{eq:CS3}
\end{equation}

Note in passing that operator products involving powers of the number operator $a^\dag_r a_r$ must first be {\it normal ordered}, i.e. all creation operators $a^\dag_r$ must preceed the annihilation operators $a_r$, before coherent states can be introduced. For example the normal order form of $\exp [a^\dag a]$ written as $:\exp [a^\dag a]:$ is expanded as
\begin{eqnarray}
:\exp [a^\dag a]: &=&  1 + (a^\dag a) + \frac{1}{2!} (a^\dag a)^2 + \cdots \nonumber \\
                              &=&  1 + (a^\dag a) + \frac{1}{2!}(a^\dag a +a^{\dag 2}a^2) + \cdots \nonumber \\
			  &=& \sum_{l=0}^{\infty} \frac{1}{l!} \sum_{j=0}^{l} s_{l,j}a^{\dag j}a^j
\label{eq:NO1}
\end{eqnarray}
where the $s_{l,j}$ are Stirling numbers of the second kind.  It follows that $:\exp [a^\dag a]:$ can be written as
\begin{equation}
:\exp [a^\dag a]: = \sum_{k=0} h_k a^{\dag k}a^k
\label{eq:NO2}
\end{equation}
\noindent where $h_k = \sum_l s_{l,k}$.

The final preliminary step in this formulation is to take the {\it continuum limit} of the expression for $\mathcal{H}$ in equation~\ref{eq:CS3}, so that
\begin{equation}
-\mathcal{H} = \int \int d^d x dt \left[ \alpha(\tilde{\vartheta} - \tilde{\varphi}) \varphi
+  (\tilde{\varphi} - \tilde{\vartheta}) \vartheta f[s(\tilde{\varphi} \varphi + \varphi)] \right]
\label{eq:CS4}
\end{equation}
where $\varphi_r \rightarrow \varphi(\mathbf{x},t) \equiv \varphi$ etc., and the conjugate coherent state $\tilde{\varphi}$ has been shifted to $1+ \tilde{\varphi}$.  

\subsection{Dimensions and the density representation}
Before proceeding further we need to assign a dimension to each variable in equation~\ref{eq:CS4}. To do so we use a modified version of the convention used in particle physics so that $[x]=L^{-1}$, $[t]=L^{-2}$ where $L$ is the length scale used, whence $[x^2/t]=L^{0}$. \{This generates a scaling found in Markov random walks and related processes such as stochastic neural activity.\} Then $[\alpha]=L^{2}$, $[\varphi]=L^{d}$, $[\tilde{\varphi}]=L^{0}$, $[\tilde{\varphi} \varphi]=L^{d}$, $[f[s]]=[\alpha]=L^{2}$. This last value of $[f[s]]$ implies that the input current function $s(\tilde{\varphi} \varphi + \varphi)=s(I)=kI$ where the constant $k$ has the dimensions of inverse current.  The net effect of such a choice leads to the required result that $[\mathcal{H}]=0$.

To emphasize this choice we further transform the coherent-state quasi-Hamiltonian by introducing the {\it density representation}:
\begin{eqnarray}
\tilde{\varphi} &\rightarrow& \exp [\tilde{n}] -1,  \varphi \rightarrow n [ \exp [ -\tilde[n] ] \nonumber \\
\tilde{\vartheta} &\rightarrow& \exp [\tilde{p}] -1,  \vartheta \rightarrow p [ \exp [ -\tilde[p] ] 
\label{eq:CS5}
\end{eqnarray}
so that equation~\ref{eq:CS4} transforms into

\begin{equation}
-\mathcal{H} = \int \int d^d x dt \left[ \alpha(\exp(\tilde{p}-\tilde{n})-1)n
+  (\exp(\tilde{n}-\tilde{p})-1) p f[s(n)] \right]
\label{eq:CS6}
\end{equation}
Note that in the continuum limit the input current function $s(n) = s(I) = kI =k(w \star n + h)$ where $\star$ is the spatial convolution operator, i.e. $w \star n=\int d^d \mathbf{x}^\prime w(\mathbf{x} - \mathbf{x}^\prime) n(\mathbf{x}^\prime, t)$.

\subsection{From the quasi-Hamiltonian to a neural Path Integral}
Using standard methods~\cite{Doi76a,Peliti85}~\citeasnoun{bc} incorporated the quasi-Hamiltonian into the action of a Wiener path integral. This action takes the form:
\begin{eqnarray}
S(n) = \int \int d^d x dt &&\left[\tilde{n} \partial_t  n + \tilde{p} \partial_t p + \right. \nonumber \\ 
&& \left. \alpha(1-\exp(-(\tilde{n}-\tilde{p}))n
-  ( \exp(\tilde{n}-\tilde{p})-1) p f[s(n)] \right]
\label{eq:Action1}
\end{eqnarray}
The utility of this action is that it is part of the exponent of the moment generating functional for the statistical moments of the probability density $P[\mathbf{n},t]$.

\noindent At an extremum 
\begin{equation}
\left.\frac{\delta S(n)}{\delta \tilde{n}} \right |_{\tilde{n}=0} = \left.\frac{\delta S(n)}{\delta \tilde{p}} \right |_{\tilde{p}=0} = 0
\label{eq:Action2}
\end{equation}
which leads to the mean-field equations
\begin{equation}
\partial_t n +\alpha n - p f[s(n)] = 0, \quad
\partial_t p -\alpha n + p f[s(n)] = 0 
\label{eq:Action3}
\end{equation}
whence
\begin{equation}
n + p = \rho
\label{eq:Action4}
\end{equation}
where $\rho$ is the (constant) packing density of excitatory neurons in the population. This is a mean-field result, so we replace $p$ by $\rho - n_{cl}$ in equation~\ref{eq:Action1}, so as to generate the mean-field Wilson-Cowan equation~\cite{Wil-Cow73} for a single spatially organized population from the first variation of the action given in equation~\ref{eq:Action1} and (in principle), all higher moments, and (finally) rewrite equation~\ref{eq:Action1} in the form:
\begin{eqnarray}
S(n) = \int \int d^d x dt &&\left[\tilde{n} \partial_t  n  + \right. \nonumber \\ 
&& \left. \alpha(1-\exp(-\tilde{n}))n
-  ( \exp(\tilde{n})-1) (\rho -n_{cl}) f[s(n)] \right]
\label{eq:Action5}
\end{eqnarray}

\subsection{Renormalizing the path integral}
We expand $n$ about its mean value $\langle n \rangle = n_{cl}$, which satisfies equation~\ref{eq:Action2} in the form:
\begin{equation}
\partial_t n_{cl} = -\alpha n_{cl} + (\rho - n_{cl} ) f[s(n_{cl} )]
\label{eq:WC}
\end{equation}
where
\begin{equation}
s(n_{cl}) = k (w \star n_{cl}  + h_{cl})
\label{eq:WC2}
\end{equation}
Thus $n \rightarrow n + n_{cl}, \tilde{n} \rightarrow \tilde{n}$, since $\tilde{n}_{cl}=0$.  So $s(n) \rightarrow s(n + n_{cl})$ and $f[s(n)] \rightarrow f[s(n + n_{cl})]$.  If follows that
$s(n+n_{cl})=k(w\star (n+n_{cl}) + (h+h_{cl}))=k(w \star n_{cl} + h_{cl}) + k(w \star n + h)) = s(n_{cl}) + s(n)$, and therefore $f[s(n)] = f[s(n_{cl}) + s(n)]$. We next expand $f[s(n)]$ in a Taylor expansion about the mean-field value $n_{cl}$, noting that from equation~\ref{eq:conv1}, $s(n)=k(w \star n + h)) =k(Ln + h)$.  In what immediately follows we assume that the external stimulus $h(\mathbf{x}, t) = 0$. It follows that:
\begin{eqnarray}
f[s(n)]=f[kL(n_{cl}+n)]=f[kLn_{cl}] &+& f^{(1)}[kLn_{cl}] kLn \nonumber \\
 &+& \frac{1}{2}f^{(2)}[kLn_{cl}] (kLn)^2 + \cdots
\label{eq:Tay1}
\end{eqnarray}
However because of normal ordering, equation~\ref{eq:Tay1} leads to the expression:
\begin{equation}
f[s(n)]=\sum_m g_m (kLn)^m, \quad \mathrm{where} \quad g_m = \sum_{l=0}^{m} \frac{f^{(l)}}{l!} s_{l,m}
\label{eq:Tay2}
\end{equation}
Since the leading terms of $g_m$ are proportional to $f^{(m)}$, and given the assumed form for $f[s(n)]$ to be such that $f^{(1)} > 0$ and $f^{(2)}<0$, then $g_m > 0$ for $m$ odd, and $g_m <0$ for $m$ even. 

\noindent We also expand the functions  $\exp(\pm \tilde{n})$. The resulting action $S(n)$ takes the form:
\begin{eqnarray}
S(n) = \int \int d^d &x&dt \left[  \tilde{n} (\partial_t + \alpha - (\rho - n_{cl}) g_1 kL) n \right. \nonumber \\
&& \left. -\frac{\tilde{n}^2}{2} (\alpha + (\rho - n_{cl}) g_1 kL )n + \tilde{n} ((\rho - n_{cl}) g_2 (kL)^2)n^2  \right. \nonumber \\
&& \left.  + \frac{\tilde{n}^2}{2} ((\rho - n_{cl}) g_2 (kL)^2)n^2  + \cdots \right] 
\label{eq:Action6}
\end{eqnarray}

\noindent It follows from the appendix that $w_2$ is small compared to $w_0$, so that in most expressions the terms proportional to $\nabla^{2m}  n^m$ can be neglected.  However this is not always the case for $m=1$.  Thus the first term can be written approximately as $\tilde{n} (\partial_t + \alpha - (\rho - n_{cl}) g_1 k(w_0 + \frac{1}{2!}w_2 \nabla^2)) n$ = $\tilde{n}(\partial_t + \mu - D \nabla^2)n$
where $\mu = \alpha -(\rho -n_{cl})g_1 k w_0$ and $D = \frac{1}{2}(\rho - n_{cl}) g_1 kw_2$. So the expression for the action is now reduced to the form:
\begin{eqnarray}
S(n) = \int \int d^d x dt &&\left[  \tilde{n} (\partial_t + \mu - D \nabla^2 ) n -\tilde{n}^2 G_1 n \right. \nonumber \\
 &&\left. + \tilde{n} G_2 n^2 + \frac{1}{2}\tilde{n}^2G_2 n^2 + \cdots \right]
\label{eq:Action7}
\end{eqnarray}
where $G_1 = 1/2 (\alpha + (\rho - n_{cl})g_1 k w_0)$, and $G_2 = (\rho - n_{cl})|g_2| k^2 w_{0}^2$.
We need to demonstrate that the last term in $S(n)$, i.e., $\frac{1}{2}\tilde{n}^2G_2 n^2$, and all other terms, are {\it irrelevant} in the sense of the renormalization group.

The renormalization group [RG] analysis is carried out via dimensional analysis.  It can be shown that all the terms in $S(n)$ are zero-dimensional when integrated over $d$-dimensional space and over time, i.e., $[d^d x dt] =L^{-(d+2)}$ and $[\mathrm{any \: term \: in \: the \: integrand}]=L^{d+2}$. However, as it stands $[n] = L^{d}$, but $[\tilde{n}] = L^{0}$, so that $[\tilde{n} n] = L^{0+d} = L^{d}$. This is not suitable for the scaling analysis implemented in the RG process.  We therefore introduce a new {\it scaling},
\begin{equation}
\tilde{s}=\sqrt{\frac{G_1}{G_2}}\tilde{n}, \quad s = \sqrt{\frac{G_2}{G_1}} n
\label{eq:S1}
\end{equation}
such that $\tilde{s}s=\tilde{n}n$ where $[G_2/G_1]=L^{-d}$.  The effect of this scaling is that both $\tilde{s}$ and $s$ have dimension $L^{d/2}$.  Let
\begin{equation}
\sqrt{G_1 G_2} = u, \quad G_2 = 2\tau
\label{eq:S2}
\end{equation}
The net effect of this scaling transformation is that
\begin{equation}
S(s) = \int \int d^d x dt \left[  \tilde{s} (\partial_t + \mu - D \nabla^2 ) s + u\tilde{s}(s-\tilde{s})s + \tau \tilde{s}^2 s^2 + \cdots \right]
\label{eq:Action8}
\end{equation}
But $[\tau/u]=L^{-d/2}$ and therefore scales to zero as $L \rightarrow \infty$ under subsequent RG transforms. So asymptotically the terms $\tau \tilde{s}^2 s^2 + \cdots$ become {\it irrelevant} in the RG sense. So finally
\begin{equation}
S(s) = \int \int d^d x dt \left[  \tilde{s} (\partial_t + \mu - D \nabla^2 ) s + u\tilde{s}(s-\tilde{s})s  \right]
\label{eq:Action9}
\end{equation}
is the {\it renormalized action} of the large-scale neural activity of a single neural population. 

\subsection{Directed Percolation}
This action is well-known: it is called Reggeon Field Theory, and is found in directed percolation [DP]  in random graphs, in contact processes, in high-energy nuclear physics, in bacterial colonies, all of which exhibit the characteristic properties of what is called a {\it universality class}, i.e., it is a phase transition with a universal scaling of important statistical exponents.  It also shows up in branching and annihilating random walks, catalytic reactions, and  interacting particles. Thus we have mapped the mathematics of large-scale neural activity in a single homogeneous neural population into a percolation problem in random graphs, or equivalently into a branching and annihilating random walk. A first version of this work was presented in~\citeasnoun{bc}. A more extensive paper with many applications to neuroscience was presented in~\citeasnoun{bc09}.

Here we note that there is an {\it upper critical dimension} at which directed percolation crosses over to mean-field behavior. This upper critical dimension is $d=4$.  What is the dimension of the neocortex? To answer this question we look at the number of synapses per neuron in the neocortex. Using estimates provided by~\citeasnoun{CS89}, this number is about $4 \times 10^3$. Assuming the number of synapses in a terminal axonal arbor to be about $50$, the number of neural neighbors per neuron is about $80$, so the effective dimensionality of a neocortical hyper lattice is about $d=40$. Thus the critical exponents characterizing the neural phase transition are the $d=4$ exponents of directed percolation. These have been calculated by~\citeasnoun{Abarbanel74}, ~\citeasnoun{Abarbanel76}, and~\citeasnoun{Amati76}, and appear in the linear response of the neocortical model to an impulsive stimulus, known to mathematicians as the {\it Green's function} and to physicists as the {\it propagator}.  This takes the form
\begin{equation}
G(x-x^\prime, t-t^\prime)\propto
\left\{
\begin{array}{cl}
	(t-t^\prime)^{-2}\exp \left[ -\frac{x-x^\prime)^2}{4(t-t^\prime)} -\mu (t-t^\prime)  \right], &  \mbox {$\mu > 0$} \\
	(t-t^\prime)^{-2}\exp \left[ -\frac{x-x^\prime)^2}{4(t-t^\prime)})\right], &  \mbox {$\mu = 0$} \\
	\mu^2 \Theta \left[ \sqrt{|\mu |}(t-t^\prime)-|x-x^\prime | \right], &  \mbox {$\mu < 0$}
\end{array}
\right.
\label{eq:Prop1}
\end{equation}
where the cases $\mu > 0, \mu = 0$, and $\mu < 0$ correspond, respectively, to the sub-critical, critical, and super-critical propagators. They correspond, respectively,  to solutions of the cable equation, the diffusion equation, and a nonlinear wave-equation.  The critical propagator is thus the diffusion limit of Brownian motion.
It turns out that there is a great deal of data supporting the hypothesis that the mean-field propagator of DP correctly describes the essential features of large-scale neocortical activity on many spatio-temporal scales. [See~\citeasnoun{Burns51},~\citeasnoun{Lampl99},~\citeasnoun{Nauhaus09}.]
In addition data on the statistical structure of large-scale activity recorded in cortical slices by~\citeasnoun{Beggs03} supports the hypothesis.  In particular, the avalanche-size distribution of spontaneous activity in cortical slices fits the DP hypothesis.

The analysis can be extended to deal with a neural network comprising both excitatory and inhibitory neurons. However in this paper we describe how to incorporate synaptic plasticity into a network comprising only excitatory neurons, as a mechanism that {\it tunes} the network so that it automatically reaches the critical point of the DP phase transition, thus exhibiting {\it self-organized criticality}.

\section{Incorporating synaptic plasticity}
Consider first a single excitatory population model with a fixed recurrent excitatory synapse $w_E$
and an input $H$ through an excitatory modifiable synapse $w_H$. 
\begin{figure}[thbp]
\begin{center}
\includegraphics[width = 9.0 cm]{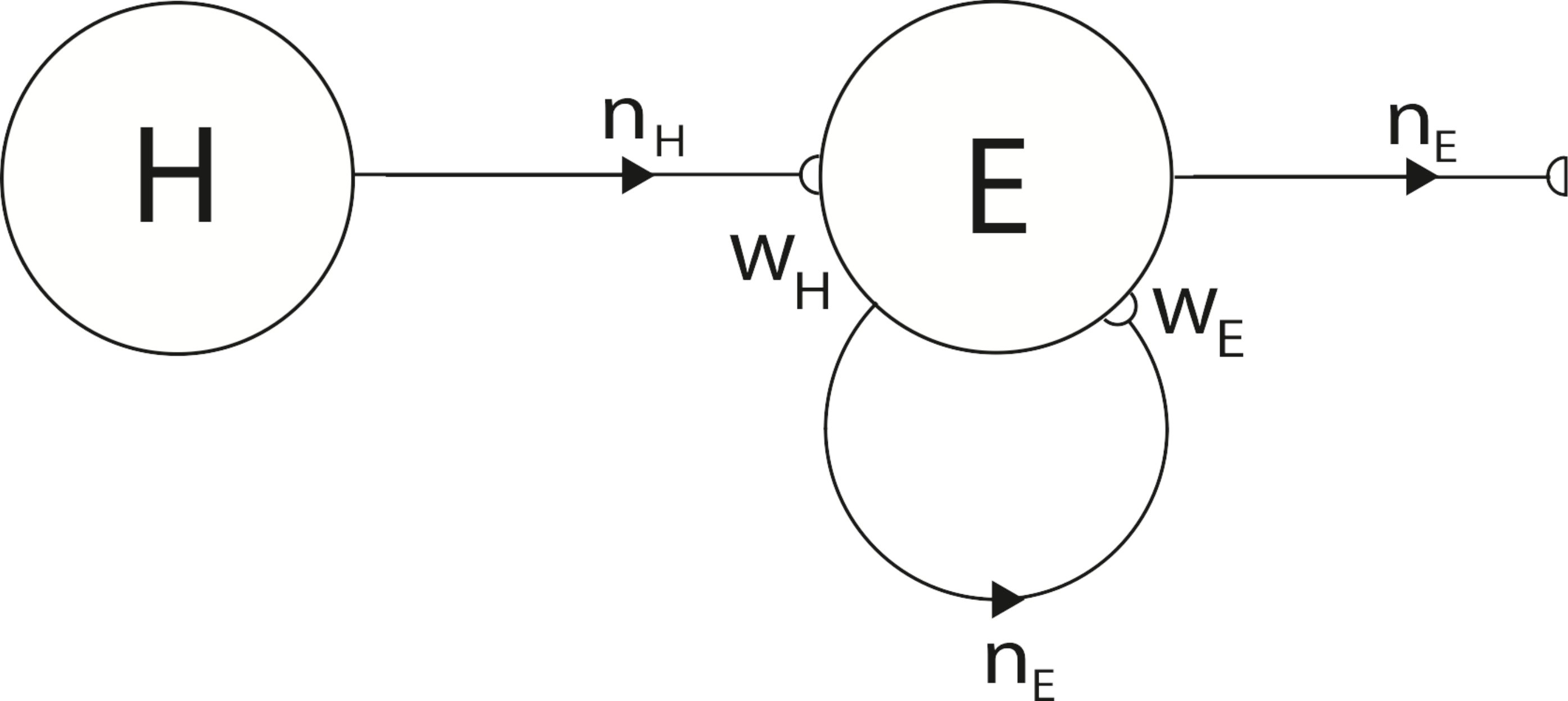}
\caption{\small A recurrent excitatory network}
\end{center}
\label{fig:F2}
\end{figure}
The mean-field neural equation for this is a version of equation~\ref{eq:WC}, i.e.
\begin{equation}
\frac{dn_E}{dt}=-\alpha_E n_E + (1-n_E) \sigma_E \left [s_E (I_E)\right]
\label{eq:neural1}
\end{equation}
where
\begin{equation}
s_E(I_E) = I_E/I_{RH,E}, \quad I_E=w_E \star n_E + w_H \star n_H
\label{eq:current}
\end{equation}
and the synapse $w_H$ is {\it modifiable} with
\begin{equation}
\frac{dw_H}{dt}=-g_E \langle \left(n_E-\rho_{E,0}-\rho_{E,S}w_H\right)n_H \rangle_t
\label{eq:plasticity1}
\end{equation}
where $\rho_{E,0}$ is a constant neural activity,  $\rho_{E,S}$ is a  constant~\cite{Vogels11}, and $g_E$ is a state-dependent 
rate function. 

\vspace{0.25in}
\noindent This is also a mean-field equation in which the synaptic weight $w_H$ is  depressed by an {\it anti-Hebbian} mechanism, and potentiated by the input activity $n_H$.  The problem is to write an action that incorporates these equations. Note that the time scale of the growth and decay of neural activity is set by the constant $\alpha_E$, whereas that of the growth and decay of synaptic plasticity is set by $g_E$.  Thus the ratio $\alpha_E/g_E$ is an important parameter.

\section{Deriving an action for synaptic plasticity}
The first problem is to develop an action for the modifiable synapse $w_H$. In order to do so we first note  from equation~\ref{eq:conv2} that $w_H$ scales with $b_H$, the synaptic conductance or {\it weight}, and we can write an approximation to eqn.~\ref{eq:plasticity1} in the form:
\begin{equation}
\frac{db_H}{dt}=-g_E \left( n_E -\rho_{E,0} -\rho_{E,S} k_H b_H \right) n_H
\label{eq:plasticity2}
\end{equation}

We next reformulate the changes in $b_H$ as a Markov process with discrete states in continuous time. We therefore assume that $b_H$ is {\it quantized} in units  $\Delta$ of synaptic weight, and similarly for $b_E$. Thus
\begin{equation}
b_E = m_E \Delta, \quad b_H = m_H \Delta
\label{eq:plasticity3}
\end{equation}
\noindent and we look at the Markov process represented in figure~\ref{fig:F3}:
\begin{figure}[thbp]
\begin{center}
\includegraphics[width = 6.0 cm]{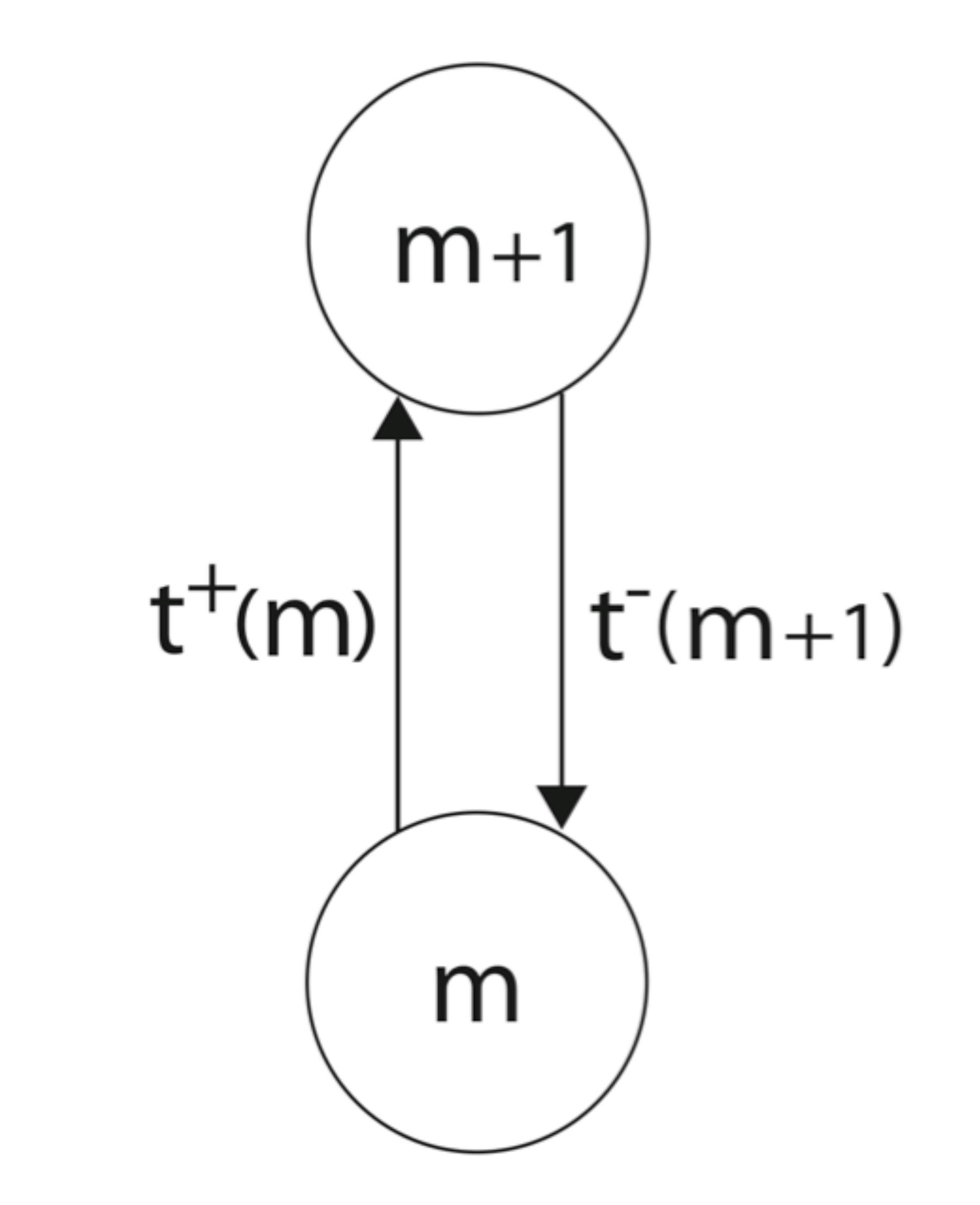}
\caption{\small Synaptic state transitions}
\end{center}
\label{fig:F3}
\end{figure}

\noindent and we write a {\it master equation} for this process in the form
\begin{eqnarray}
\frac{dP(m_H,t)}{dt}&=&t^{-} (m_H +1)P(m_H +1,t)-t^{-}(m_H)P(m_H,t) \nonumber \\
&+&t^{+} (m_H -1)P(m_H -1,t)-t^{+}(m_H)P(m_H,t)
 \label{eq:ME9}
\end{eqnarray}
where $P(m, t)$ is the probability that the synaptic weight  $m_H = m$ at time $t$.
and
\begin{eqnarray}
t^{+}&=&g_E (\rho_{E,0}+\rho_{E,S}k_H m_H)n_H \nonumber \\
t^{-}&=&g_E n_E n_H
\label{eq:ME10}
\end{eqnarray}
are the transition rates for the excitatory synapse $m_H$.

\subsection{The van Kampen ladder operators}
We again introduce the van Kampen ladder operators
\begin{equation}
E^{\pm}_{m} m = m \pm 1
\label{eq:ME11}
\end{equation}
so that the master equation can be rewritten as:
\begin{equation}
\frac{dP(m_H,t)}{dt}=\left[t^{-}(E^{+}_{m_H} -1)+t^{+}(E^{-}_{m_H}-1)\right]P(m_H,t)
\label{eq:ME12}
\end{equation}

An examination of eqns.~\ref{eq:ME9}-\ref{eq:ME11} indicates that only the transition rate $t^{+}$ contains a term proportional to $m_H$.  We will utilize this property in what follows.

\subsection{From the Master Equation to the Action}
We now introduce bosonic annihilation and creation operators for $m_H$. Let such operators be denoted by $s^{\dag}$ and $s$ respectively, and let $ |m \rangle$ be a column vector representing the synaptic weight $m$ such that 
\begin{equation}
s^{\dag} |m \rangle = | m + 1\rangle, \quad s |m \rangle = m |m -1 \rangle
\label{eq:B1}
\end{equation} 

\noindent Note that $s^{\dag}$ and $s$ satisfy the same commutation rules and equations that we introduced earlier, so that
\begin{eqnarray}
E^{+}_{m}-1 &\rightarrow& s^{\dag} - s^{\dag}s = s^{\dag}(1-s) \nonumber \\
E^{-}_{m}-1 &\rightarrow& s- s^{\dag}s = (1 - s^{\dag})s
\label{eq:B5}
\end{eqnarray}
\noindent So we consider the equation:
\begin{equation}
\frac{\partial |\hat {m} \rangle}{\partial t} =(E^{-}_{m}-1)|m\rangle = (1-s^{\dag}) s |m \rangle
 \label{eq:Map}
\end{equation}

\noindent We further note that $(1-s^\dag )s$ is {\it normal ordered}. We therefore shift $s^\dag$ and $s$,
\begin{equation}
s^\dag \rightarrow 1+ \tilde{s}, s \rightarrow s
\label{eq:CS7}
\end{equation}
where $\tilde{s}$ and $s$ are now {\it coherent states}, and introduce the {\it density representation}
\begin{equation}
\tilde{s} \rightarrow \exp(\tilde{m})-1, \quad s \rightarrow m \exp(-\tilde{m})
\label{eq:CS8}
\end{equation}
\noindent we find
\begin{eqnarray}
\partial_t |m\rangle&=&-\tilde{s}s |m\rangle \nonumber \\
&=& -(\exp(\tilde{m})-1)m\exp(-\tilde{m})|m\rangle \nonumber \\
&=& -(1-\exp(-\tilde{m}))m|m\rangle
\label{eq:CS9}
\end{eqnarray}

\noindent Eqn.~\ref{eq:CS3} tells us that the action $S_{m}$ must contain a term of the form
$$-g_E \rho_{E,S} k_H n_H (1-\exp(-\tilde{m}_H))m_H,$$ a {\it source} term $$-g_E \rho_{E,0} n_H \tilde{m}_H,$$ and an {\it interaction} term of the form $$+g_E  n_E n_H \tilde{m}_H,$$
leading to an action of the form:
\begin{eqnarray}
S(m_H)=\int \int d^d x dt &&\left[ \tilde{m}_H \partial_t m_H - g_E \rho_{E,S} k_H n_H (1-\exp(-\tilde{m}_H))m_H \right. \nonumber \\
&&\left.-\tilde{m}_H g_E \rho_{E,0} n_H + \tilde{m}_H g_E n_E n_H \right]
\label{eq:Action10}
\end{eqnarray}

\noindent Using variational techniques, we can derive the mean-field equation [eqn~\ref{eq:plasticity1}] from the condition:
\begin{equation}
\left.\frac{\delta S(\tilde{m}_H)}{\delta \tilde{m}_H} \right |_{\tilde{m}_H=0} = 0
\label{eq:plasticity4}
\end{equation}
\noindent For then we obtain the equation:
\begin{equation}
\frac{dm_H}{dt}=-g_E \left(n_E - \rho_{E,0} -\rho_{E,S} k_H m_H \right) n_H
\label{eq:plasticity5}
\end{equation}
 i.e., eqn.~\ref{eq:plasticity2}.
 
 \subsection{Renormalizing the synaptic plasticity action}
We now proceed to renormalize the action $S(m_H)$ just as we renormalized $S(n_E)$.
 We therefore expand the exponential term in equation~\ref{eq:Action10} and rewrite $S(m_H)$ in the form: 
 \begin{eqnarray}
S(m_H)=\int \int d^d x dt &&\left[ \tilde{m}_H \partial_t m_H - H_1 \tilde{m}_H m_H n_H +\frac{1}{2} H_1 \tilde{m}^2_H m_H n_H  \right. \nonumber \\
&&\left.- H_2 \tilde{m}_H  n_H + \tilde{m}_H g_E n_E n_H \right]
\label{eq:Action11}
\end{eqnarray}
where $H_1 =  g_E \rho_{E,S} k_H$, and $H_2 =g_E \rho_{E,0}$.

\noindent We now introduce the scaling
\begin{equation}
\tilde{s}_H = \sqrt{\frac{H_2}{H_1}}\tilde{m}_H, \quad {s}_H = \sqrt{\frac{H_1}{H_2}}{m}_H
\label{eq:S3}
\end{equation}
such that $\tilde{s}_H s_H - \tilde{m}_H m_{H}$, and $[H_1/H_2]=L^{-d} $.   This scaling is analogous to the scaling of $n$ and $\tilde{n}$ which we carried out earlier for neural activities.  As before the effect of this scaling is that both $\tilde{s}_H$ and $s_H$ have dimension $L^{d/2} $.

\noindent Let
\begin{equation}
\sqrt{H_1 H_2} = u_H, \quad H_1 = 2 \tau_H
\label{eq:S4}
\end{equation}
and recall that equation~\ref{eq:S1} scales $n_E$ to $\sqrt{{G_1}/{G_2}} s_E$.

 Following the procedure outlined earlier we can calculate which terms in the transformed action $S(s_H)$ become irrelevant under scaling transformations. The resulting renormalized synaptic plasticity action takes the form:
\begin{equation}
S(s_H) = \int \int d^d x dt \left[ \tilde{s}_H \partial_t s_H - u_H \tilde{s}_H n_H \right]
\label{eq:Action12}
\end{equation}

\section{Combining the actions}
It follows from this formulation that the full action for the coupled system of equations for the evolution of $n_E$ and $m_H$ can be obtained simply by adding the actions $S(n_E)$ and $S(m_H)$ together.  The combined action therefore takes the form:
\begin{eqnarray}
S(n_E , m_H) = \int \int && d^d x dt \left[\tilde{n}_E \partial_t  n_E  + \alpha(1-\exp(-\tilde{n}_E))n_E \right. \nonumber \\ 
&& \left.  -  ( \exp(\tilde{n}_E)-1) (\rho -n_{E,cl}) f[s(n_E)] + \tilde{m}_H \partial_t m_H \right. \nonumber \\
&&\left.  - g_E \rho_{E,S} k_H n_H (1-\exp(-\tilde{m}_H))m_H \right. \nonumber \\
&&\left.-\tilde{m}_H g_E \rho_{E,0} n_H + \tilde{m}_H g_E n_E n_H \right]
\label{eq:Action13}
\end{eqnarray}

\subsection{A simulation of the behavior of the combined mean-field equations}
The first variation of equation~\ref{eq:Action13}  generates the mean-field equations for $n_E$ and $m_H$in the form:
\begin{eqnarray}
\frac{dn_E}{dt}&=&-\alpha_E n_E + (1-n_E) \sigma_E \left [s_E (I_E)\right] \nonumber \\
\frac{dm_H}{dt}&=&-g_E \left( n_E -\rho_{E,0} -\rho_{E,S} k_H m_H \right) n_H 
\label{eq:comb1}
\end{eqnarray}
\noindent 
where
\begin{equation}
s_E(I_E)=k_E (m_E \star n_E + m_H \star n_H)  
\label{eq:current2}
\end{equation}

\noindent These equations can be simulated.  The results are shown in figure 4.
\begin{figure}[thbp]
\begin{center}
\includegraphics[width = 12.0 cm]{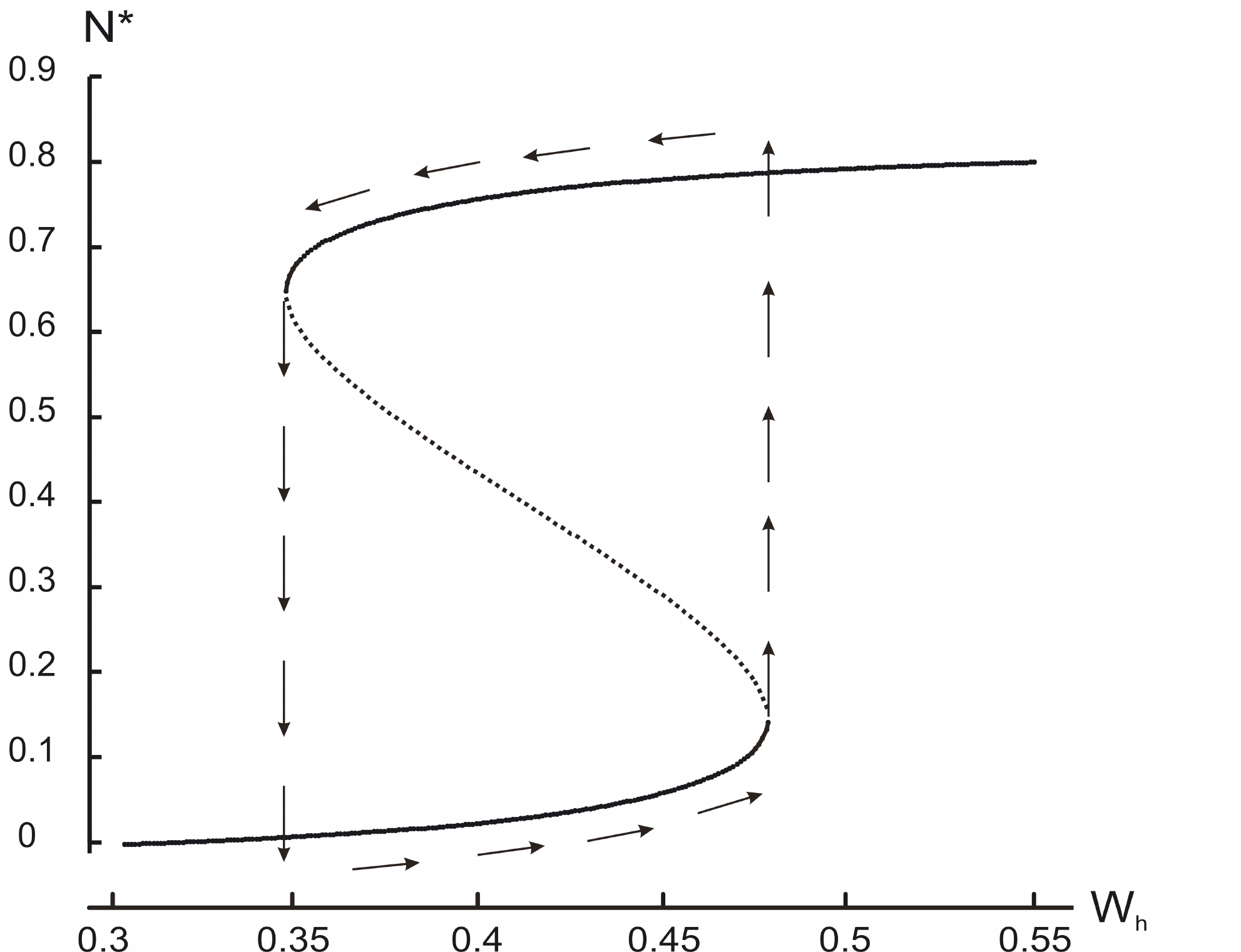}
\caption{\small Neural state transitions between a ground state and an excited state. Parameter values: $m_E = 3, n_H = 3; \alpha = 0.2$. $N^\star$ is the fixed-point value of $n_E$, and $W_H$ is the magnitude of the anti-Hebbian synapse in the input path.}
\end{center}
\label{fig:F4}
\end{figure}
It will be seen that in the ``ground-state'' of low values of $N^\star = n_E^\star$ the synaptic weight $w_H$ (proportional to $m_H$), increases until it reaches the critical point at a saddle-node bifurcation, at which point $N^\star$ becomes unstable and the system switches to the ``excited-state''. But then the anti-Hebbian term in the synaptic plasticity dynamics kicks in, and $W_H$ declines until the excited-state fixed-point becomes unstable at the upper critical point, also at a saddle-node, and switches back to the ground-state fixed point, following which the hysteresis cycle starts over. This is a exact representation of the sand-pile model's behavior.  This is another representation of SOC in a neural network. The reader should compare this with the synaptic mechanisms for achieving SOC described in~\citeasnoun{Lev07} and in~\citeasnoun{Nie10}.

\subsection{Renormalizing the combined action}
To renormalize the combined action we follow the same procedure as before and expand the exponential functions $\exp[ \pm \tilde{n}_E ]$, $\exp[ \pm \tilde{m}_H ]$, and  $f[s(n_E)]$. Note that after normal ordering and collecting terms, $f[s(n_E)]$ can be expanded in the extended form
$f[s(n_E)] = \sum_m g_m (k_E (L_E n_E+L_H n_H))^m$. After scaling and dimensional analysis, the renormalized combined action takes the form:
\begin{eqnarray}
S(s_E, s_H)=\int \int d^d x dt && \left[  \tilde{s}_E (\partial_t + \mu_E - D_E \nabla^2 ) s_E + u_E \tilde{s}_E (s_E -\tilde{s}_E)s_E \right. \nonumber \\
&& \left. -v_E \tilde{s}_E n_H + \tilde{s}_H \partial_t s_H - u_H \tilde{s}_H n_H \right]
\label{eq:Action14}
\end{eqnarray}
where $u_E, u_H$ and $v_E$ are renormalized constants. It would appear  that apart from the source term $v_E \tilde{s}_E n_H$ the coupled action is just the sum of the two uncoupled renormalized actions.   This is indeed the case!  All the addition terms which appear in the current function become irrelevant under renormalization, and do not effect the renormalized action representing $n_E$. Similarly for $m_H$.  It follows that fluctuations in the activity $n_E$ in the neighborhood of the critical point $\mu_E = 0$, i.e. those in the fluctuation-driven regime, should be essentially those of directed percolation.

\section{Inhibitory synapses}
In case $w_H$ is an {\it inhibitory} synapse the transitions $t^{+}(m)$ and $t^{-}(m)$
are reversed. so that:
\begin{eqnarray}
t^{+}&=&g_E  n_E n_H  \nonumber \\
t^{-}&=&g_E (\rho_{E,0}+\rho_{E,S}k_H m_H)n_H
\label{eq:ME13}
\end{eqnarray}

\noindent Thus now only the transition rate $t^{-}$ contains a term proportional to $m_H$.
The effect of this is that the action for $S_{m}$ leads to the mean-field equation:
\begin{equation}
\frac{dm_H}{dt}=g_E \left(n_E - \rho_{E,0} -\rho_{E,S} k_H m_H \right) n_H
\label{eq:plasticity6}
\end{equation}
at an inhibitory synapse $m_H$.

Thus inhibitory feedforward synapses are Hebbian with stimulus dependent depression, whereas excitatory feedforward synapses are anti-Hebbian with stimulus dependent potentiation,  and we have now formulated actions for feedforward excitatory and inhibitory synaptic plasticity, based on simple microscopic potentiation and depression processes, involving a unary variable, the synaptic weight $m_H$.  

\section{Conclusion}
In this paper we have indicated how one can formulate and analyze actions for a simple network of excitatory cells with an input coupled to the network via an activity-dependent modifiable synapse.  This action allows, in principal, the computation of statistical moments of the fluctuating dynamics of the network.   Previous work by~\citeasnoun{Wil-Cow72} indicates that the mean-field dynamics of the network is bistable, and can generate a hysteresis loop. On coupling this dynamics to that of the modifiable feedforward synapse introduced, all the conditions for the achievement of SOC are present in the combined system.  The simulation of the mean-field dynamics shows that SOC is indeed achieved.  It only remains to simulate the behavior of the combined system with intrinsic noise.  Our prediction is that the fluctuations in the activity near the critical points of the system will exhibit the properties of directed percolation. This will be the subject of another paper.

\ack
The work reported in this paper was initially developed in large part with Michael. A. Buice. The current work was supported (in part) by a grant to Wim van Drongelen from the Dr. Ralph \& Marian Falk Medical Trust.

\appendix
\section*{Appendix}
\setcounter{section}{1}
\subsection{Expanding the weighting function}
We approximate the convolution $w \star n$  as:
\begin{eqnarray}
w \star n &=& \int d^d x^\prime w (\mathbf{x}-\mathbf{x}^\prime) n (\mathbf{x}^\prime ,t) \nonumber \\
&\simeq& (w_0 + \frac{1}{2!}w_2 \nabla^2 + \cdots) n (\mathbf{x}^\prime ,t) \nonumber \\
&\equiv& L n
\label{eq:conv1}
\end{eqnarray}
where 
\begin{equation}
 w (\underline{x}) \rightarrow w (r) = b/\sigma^d e^{-r/\sigma}, \sigma =r_0.
 \label{eq:conv2}
 \end{equation}
It  follows that
\begin{eqnarray}
w_0&=&\int d^d x w(\mathbf{x}) = b  \frac{ d\Gamma (d) }{\Gamma(d/2 + 1)}\pi^{d/2},  \nonumber \\
w_2&=&\int d^d x^2 w(\mathbf{x}) = b \sigma^2  \frac{ d\Gamma (d+2) }{\Gamma(d/2 + 1)}\pi^{d/2}
\label{eq:conv3}
\end{eqnarray}
In case $d = 3$, $w_0 = 8 \pi b, w_2=12w_0 \sigma^2$, whence  $w_2/w_0 = 12\sigma^2$. 		
This expansion of the weighting function $w(\mathbf{x})$ is known as the {\it moment} expansion.

\section*{References}
\bibliographystyle{jphysicsB}

\bibliography{Refs10}

\end{document}